\documentclass[singlecolumn, superscriptaddress,showpacs, secnumarabic,amssymb,nobibnotes,aps,pre]{revtex4-2}
\usepackage{graphicx, graphics, amsmath, amssymb, rotate, textcomp, gensymb}
\usepackage{mathrsfs, float}
\usepackage[T1]{fontenc}
\usepackage{dcolumn}
\usepackage{physics}
\usepackage{wrapfig}
\usepackage{xcolor}
\usepackage[none]{hyphenat}
\usepackage{url,hyperref}
\usepackage{times}
\usepackage{multirow}
\begin{document}

\title{Density-field structures in a few systems undergoing velocity ordering}

\author{Subir K. Das}
\email{Email of corresponding author: das@jncasr.ac.in}
\affiliation{Theoretical Sciences Unit and School of Advanced Materials, Jawaharlal Nehru Centre for Advanced Scientific Research, Jakkur, Bangalore 560064, India}
\author{Sohini Chatterjee}
\affiliation{Theoretical Sciences Unit and School of Advanced Materials, Jawaharlal Nehru Centre for Advanced Scientific Research, Jakkur, Bangalore 560064, India}
\author{Wasim Akram}
\affiliation{Theoretical Sciences Unit and School of Advanced Materials, Jawaharlal Nehru Centre for Advanced Scientific Research, Jakkur, Bangalore 560064, India}
\author{Subhajit Paul}
\affiliation{Theoretical Sciences Unit and School of Advanced Materials, Jawaharlal Nehru Centre for Advanced Scientific Research, Jakkur, Bangalore 560064, India}
\affiliation{Department of Physics and Astrophysics, University of Delhi, Delhi 110007, India}
\author{Saikat Chakraborty}
\affiliation{Theoretical Sciences Unit and School of Advanced Materials, Jawaharlal Nehru Centre for Advanced Scientific Research, Jakkur, Bangalore 560064, India}
\affiliation{Le laboratoire interdisciplinaire de Physique, Université Grenoble Alpes, 140, Rue de la Physique, 38402 St. Martin d’Hères, France.}
\author{Arabinda Bera}
\affiliation{Theoretical Sciences Unit and School of Advanced Materials, Jawaharlal Nehru Centre for Advanced Scientific Research, Jakkur, Bangalore 560064, India}
\affiliation{Department of Physics “A. Pontremoli”, University of Milan, via Celoria 16, 20133 Milan, Italy}

\date{\today}

\begin{abstract}
We consider two (off-lattice) varieties of out-of-equilibrium systems, viz., granular and active matter systems, that, in addition to displaying velocity ordering, exhibit fascinating pattern formation in the density field, similar to those during vapor-liquid phase transitions. 
In the granular system, velocity ordering occurs due to reduction in the normal components of velocities, arising from inelastic collisions. 
 In the active matter case, on the other hand, velocity alignment occurs because of the inherent tendency of the active particles to follow each other. 
 Inspite of this difference, the patterns, even during density-field evolutions, in these systems can be remarkably similar. This we have quantified via the calculations of the two-point equal time correlation functions and the structure factors. These results have been compared with the well studied case of kinetics of phase separation within the framework of the Ising model. Despite the order-parameter conservation constraint in all the cases, in the density field, the quantitative structural features in the Ising case is quite different from those for the granular and active matters. Interestingly, the correlation function for the latter varieties, particularly for an active matter model, quite accurately describes the structure in a real assembly of biologically active particles.
\end{abstract}

\maketitle

\section{Introduction}
\sloppy

Starting from the structure in a bacterial colony to the arrangement of galaxies, stars and planets in the universe, pattern formation is ubiquitous and occurs at all known length scales \cite{cross_rmp,bray199341}.  
Despite the differences in interactions, among the constituents, striking resemblance is often observed in patterns appearing in diverse varieties of systems \cite{cross_rmp,bray199341}. 
Homogeneous granular systems, having constituents that undergo inelastic collisions, develop interesting structures in velocity, as well as in the density fields, as the systems cool \cite{goldhirsch1993clustering,paul_ba_2017,das2003}. The particles in this example do not self-propel, \textit{i.e.}, the systems are passive. Nevertheless, the observed patterns can interestingly be similar to those in certain systems with self-propelling or active constituents \cite{Vicsek1995}. In such active matter systems, a realistic way to incorporate activity is through alignment of a particle's velocity  along that of its neighbors. 
This protocol constitutes the Vicsek model that is widely used as the (useful) minimal model to study pattern formations in a class of living matters~\cite{Vicsek1995,chate2020dry}. 
Such activity rule, when combined with environmental noise, can lead to order-disorder transitions, associated with regions of high and low densities of particles, resembling vapor-liquid phase separation in a standard condensed matter system~\cite{cross_rmp,landau_binder,bray199341}. Both these features are observed in granular media as well. Here note that, self-propulsion, reason for the pattern formation, in active particles is generated by continuous supply of energy. The passive (here, granular) systems can exhibit similar phenomena even in absence of such supply. 

Despite apparent similarities in structures in the two cases, a comparative study is missing at the quantitative level, to the best of our knowledge. Using molecular dynamics (MD) simulations \cite{allen2017} and other numerical (update) rules \cite{Vicsek1995,Das2017,Chakraborty2020,Paul2021,paul2024finite}, we study here the patterns developed in a freely cooling granular gas (FCGG) and aligning active matters. 
Given that for all of these the density field order-parameter, during phase separation, remains conserved, we are motivated also to compare the structural aspects of these model systems with those for the conserved Ising model (CIM) \cite{landau_binder} for which we carry out Monte Carlo (MC) simulations ~\cite{landau_binder}.   
To quantify the (density-field) structures, we calculate the two-point equal time correlation function \cite{bray199341}, for (expected) isotropic patterns, as 
\begin{equation}
  C(r,t)= \langle \psi(\vec{r},t) \psi(0,t) \rangle - \langle \psi(\vec{r},t) \rangle \langle \psi(0,t) \rangle.  
\end{equation}
Here, $\psi(\vec{r},t)$ is the relevant (time-dependent) local order parameter, $r~(=|\vec{r}|)$ is the scalar distance between two points and $\langle ... \rangle$ indicates circular averaging as well as that over different initial configurations. The self-similarity of the growth process requires scaling of the correlation function as \cite{bray199341}     
\begin{equation}
C(r,t) \equiv \tilde{C}(r/\ell(t)),
\end{equation}
where $\ell(t)$ is the characteristic length-scale of the system, $\tilde{C}$ being a time-independent function. When such scaling is obeyed, it is understood that two structures differ from each other only by the size $\ell$ .
We also calculate the structure factor, $S(k,t)$, the Fourier transform of $C(r,t)$ \cite{bray199341}, $k$ being the magnitude of the wave vector, a quantity that is directly measurable in experiments.

\section{Model and Methods}

For the granular system, the particles (of finite diameter $\sigma$) are of hard-sphere nature. Thus, we perform event-driven MD simulation \cite{allen2017} during which we keep track of collision times and the colliding partners. The post-collisional velocities of the partners $i$ and $j$, for a coefficient of restitution $e~(<1)$, are updated as \cite{goldhirsch1993clustering,das2003,paul_ba_2017} 
\begin{equation}
\begin{split}
\vec{v}~'_{i} = \vec{v}_{i} - (1+e)(\vec{v}_{ij}\cdot \hat{n})~\hat{n}/2, \\
\vec{v}~'_{j} = \vec{v}_{j} + (1+e)(\vec{v}_{ij}\cdot \hat{n})~\hat{n}/2,    
\end{split}
\end{equation}
where $\vec{v}_{i}$ and $\vec{v}_{j}$ are pre-collisional velocities, and corresponding primed quantities are the post-collisional ones. Here $\vec{v}_{ij}
~(=\vec{v}_i - \vec{v}_j)$ is the relative velocity and  $\hat{n}~(= (\vec{r}_i-\vec{r}_j)/|\vec{r}_i-\vec{r}_j|)$ defines the direction of the unit vector joining the particles $i$ and $j$. Such an athermal system cools freely, in absence of any external drive, the degree of energy dissipation being set by $e$.  
We label the standard Vicsek model \cite{Vicsek1995} as Active Matter Model $1$ (AMM$1$). Here, the $j$ th particle, moving with velocity  $\vec{v}_j = v_0 e^{i \theta_j}$, at time $t$, updates its direction $\theta_j$, after a time step $\Delta t$, in such a way that the new direction tends to align with the average orientation of neighboring particles, subject to a perturbation $\delta \theta_j$, the latter being modeled as random noise. The corresponding update rule is given by~\cite{Vicsek1995} 
\begin{equation}
\theta_j(t+\Delta t) = {\rm arg} {[ \sum_{k=1}^{N_{R_c}} e^{i \theta_k (t)}]} +\delta \theta_j,
\end{equation}
where $N_{R_c}$ represents the number of neighboring particles within a cut-off radius $R_c=2.0$. The value of $\delta \theta_j$, at each step, is randomly chosen from a stochastic noise within a range $[-\eta/2, \eta/2]$  (noise strength $\eta \in [0,2\pi]$). 
The position of the particle, $\vec{r}_j$, is updated for time $t+\Delta t$ (with $\Delta t=1$) as 
\begin{equation}
\vec{r}_j (t+\Delta t) = \vec{r}_j (t) + \vec{v}_j (t+\Delta t) \Delta t.
\end{equation}
In the other model of active matter \cite{Das2017}, to be referred to as AMM$2$, both inter-particle passive interaction and Vicsek-like self-propulsion are included \cite{Chakraborty2020,Paul2021,Das2017,paul2024finite}. 
In presence of a passive interaction potential $U_i$, modeled via the Lennard-Jones pair interaction \cite{allen2017} 
\begin{equation}
V_{LJ}= 4\epsilon \left[\left({\sigma}/{r}\right)^{12}- \left({\sigma}/{r}\right)^{6}\right]
\end{equation}
(for modification due to truncation see Ref. \cite{Das2017}), the equation of motion for the $i$-th particle we write as \cite{Das2017} 
\begin{equation}
\frac{d^2\vec{r}_i}{dt^2} = -\gamma \frac{d\vec{r}_i}{dt}- \vec{\nabla} U_i+ \sqrt{2\gamma k_BT} \eta_i(t)+f_A \hat{n}_i,
\end{equation}
$\eta_i$ being $\delta$-correlated Gaussian white noise with zero mean and unit variance for which the two time correlation is defined as 
\begin{equation}
\langle \eta_i(t)\eta_i(t')\rangle=\delta(t-t'). 
\end{equation}
We take $0.2$ as the value of temperature ($T$), in appropriate LJ unit \cite{Das2017,Chakraborty2020,Paul2021,paul2024finite}. Values of all the other parameters, viz., interaction strength ($\epsilon$), particle diameter ($\sigma$) and damping constant ($\gamma$),  are set at $1$.  The integrations are performed using Verlet velocity algorithm \cite{allen2017}, with time step $\Delta t=0.002$, in an LJ unit. The velocities of the particles ($\vec{v}_i^{\text{pas}}$) are first obtained by solving the above Langevin equation \cite{Das2017,Chakraborty2020,Paul2021,paul2024finite}, without the activity term $f_A \hat{n}_i$. Then the final velocity ($\vec{v}_i^f$) has been obtained via the vector sum \cite{paul2024finite} of $\vec{v}_i^{\text{pas}}$ and $\Delta t f_A \hat{n}_i$, with $\hat{n}_i$ being the average direction of the neighbors, lying within $R_c=2.5\sigma$. This operation is carried out in such a way that the direction of velocity only changes \cite{Das2017,Chakraborty2020,Paul2021,paul2024finite}. For the activity strength, analogous to the inverse of $\eta$ in AMM1, we choose $f_A=1.0$, unless otherwise mentioned.  Contrary to the AMM1, in which interaction solely arises from the dynamic alignment rule, here due to the additional inter-particle LJ interaction, the density within the clustered phase cannot attain a very high value. For all the models, the simulations are performed with (number) density $\rho \simeq 0.37$, starting with particles randomly placed in (periodic) square boxes of linear dimension $L=256$, with uncorrelated velocities.

\section{Results}

\begin{figure}[ht!]
\centering
\includegraphics[width=8cm,clip]{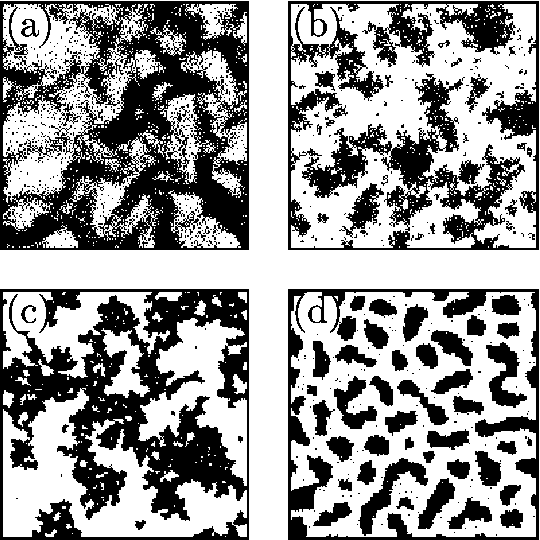}
\caption{Representative snapshots recorded during evolutions are shown for (a)~FCGG ($t=10^2$), (b) AMM$1$ ($t=10^3$), (c) AMM$2$ ($t=10^2$) and (d) CIM ($t=10^5$). In (a)-(c) black dots mark the positions of the particles, whereas in (d) black dots indicate the lattice points occupied by $+1$ spins.}
\label{fig-1}       
\end{figure}

We start by showing representative snapshots of the patterns developed in  FCGG, AMM$1$ and AMM$2$, in Fig.~\ref{fig-1}(a-c).
(For the active matter cases the homogeneous systems were quenched deep inside the ordered regions of the phase diagram, with the variations of $\eta$ or $f_A$.) 
 Similar morphologies, with domains of condensed and dilute phases, can be clearly observed. 
Fig.~\ref{fig-1}(d) shows a snapshot during evolution in the \rm{2D} CIM. 
The corresponding Hamiltonian is \cite{landau_binder}
\begin{equation}
    H = -J\sum_{\langle ij\rangle} S_i S_j, 
\end{equation}
$\langle ij \rangle$ defining the nearest-neighbor summation. For the interaction strength $J$, we choose $J=+1$, 
implying  ferromagnetic ordering \cite{landau_binder}. The Hamiltonian represents a two-component system, given that $S_i$ (located on grids of a square lattice) can take values $+1$ or $-1$, representing $A$ and $B$ particles, respectively, in a binary mixture. To maintain consistency, here also we keep the density of occupied points as $0.37$. In this  case, systems, with random initial arrangements of $A$ and $B$ particles are probed following sudden quenches to a final temperature $T=0.6T_c$, where $T_c(\simeq 2.27 J/k_B$, $k_B$ being the Boltzmann constant) is the critical temperature \cite{landau_binder}.
The developed structures, resembling somewhat elongated droplets, are significantly different from the other three cases. For this case, the unit of time is an MC step \cite{landau_binder}. 

\begin{figure*}[ht!]
\centering
\includegraphics[width=18cm, clip]{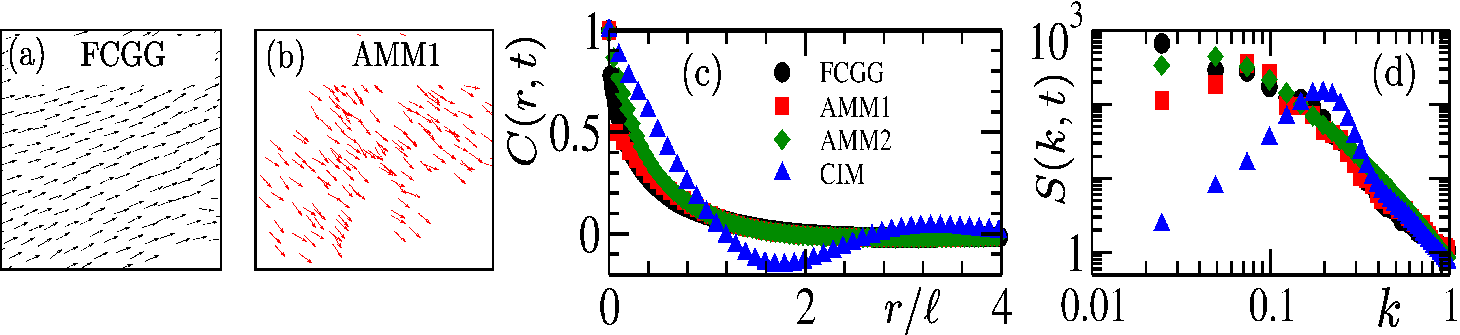}
\caption{(a) Velocity field ordering in a high density region during the evolution in a FCGG. (b) Similar to (a), but here we demonstrate velocity ordering for the AMM$1$. (c) Plot of the density-field correlation functions, $C(r,t)$, versus $r/\ell$, for all the models: FCGG ($t=10^2$), AMM$1$ ($t=10^3$), AMM$2$ ($t=10^2$) and CIM ($t=10^5$). In each of the cases the abscissa has been scaled by the values of $\ell$, calculated as $C(r=\ell)=0.1$. (d) Plots of the structure factor $S(k,t)$, versus $k$, for FCGG, AMM1, AMM$2$, and the CIM, from times quoted in (c).}
\label{fig-2}       
\end{figure*}

While all the four systems follow conserved dynamics \cite{bray199341}, during phase separation, the patterns appear to be similar for systems undergoing velocity ordering.
Note that for the CIM there is no associated velocity field.
In Fig.~\ref{fig-2}(a,b), we show velocity ordering within the dense phases of the FCGG and AMM$1$. In the case of the FCGG, due to inelastic collisions, the normal components of velocities of colliding particles decrease, the tangential components remaining unchanged, leading to the velocity parallelization.
However, in the active case, the dynamical rule of local velocity alignment governs the ordering behavior. 
Furthermore, while the velocities of the particles keep decreasing in the granular case, due to the lack of an external energy input, velocities in active matter flock can be maintained due to self-propulsion via inherent energy pumping. 
In spite of the different underlying kinetics of velocity ordering, the resultant density fields appear to be sufficiently similar.  

To quantitatively verify the similarities, we probe the patterns \textit{via} calculations of $C(r,t)$. 
In Fig.~\ref{fig-2}(c) we plot the results for all the considered systems. For convenience, the abscissa is scaled with $\ell$, the average domain size. 
For the cases having velocity ordering, viz., FCGG, AMM$1$ and AMM$2$, reasonably good collapse of data can be appreciated. 
In agreement with a discussion above, the $C(r,t)$ for the Ising model differs significantly from the rest. In this case, $C(r,t)$ exhibits strong oscillatory behavior which can be attributed to the quasi-periodic density field -- see Fig. \ref{fig-1}(d). 

As an alternative, in Fig.~\ref{fig-2}(d), on a double-log scale, we plot the $S(k,t)$, to a multiplicative factor, for the representative systems in Fig.~\ref{fig-2}(c). 
The interfacial properties are captured in the decay at large $k$, which appears to be (expectedly) similar in all the cases. 
Furthermore, for the systems exhibiting velocity ordering, we observe nice agreement of data for $S(k,t)$ across scales, apart from slight mismatch in the small $k$ regime.  
The $S(k,t)$ for the CIM, however, shows contrasting behavior at small $k$.

\begin{figure}[!ht]
\centering
\includegraphics[width=9cm,clip]{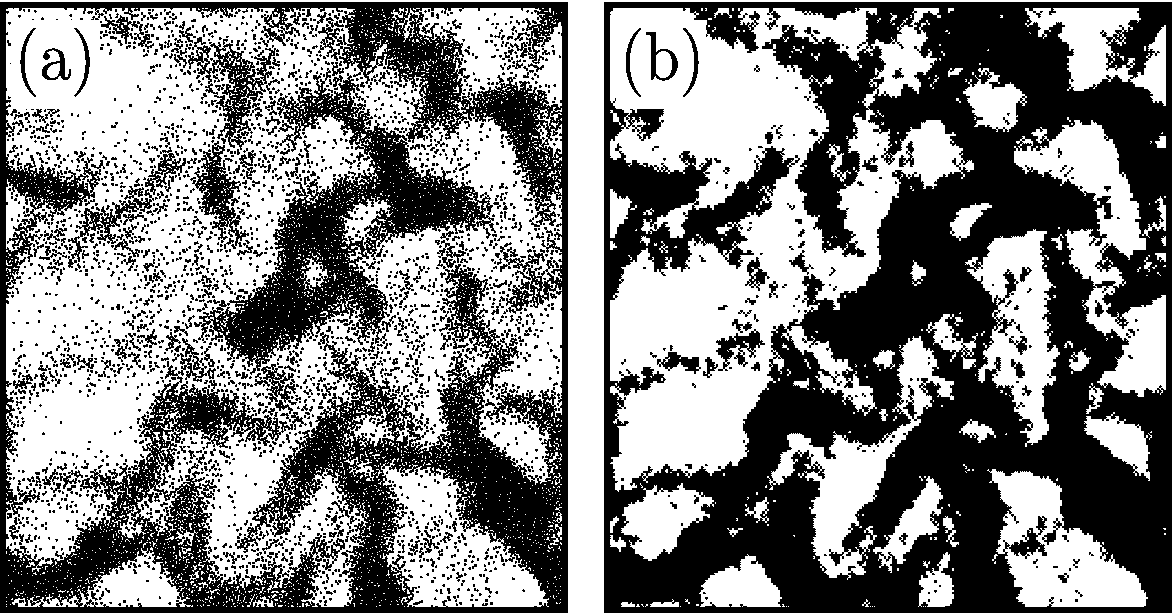}
\caption{(a) A snapshot obtained during an evolution of the FCGG. (b) Corresponding coarse-grained picture.}
\label{fig-3}       
\end{figure}

Next we investigate how these models can reproduce structures seen in real systems. For the latter we choose a school of fish.
Often such systems are contaminated with noisy backgrounds, which hinder accurate calculation of the structural quantities. 
To circumvent the difficulty, it is necessary to coarse-grain the systems, following a majority population rule \cite{paul_ba_2017}, to capture the basic structural feature. This is demonstrated in Fig.~\ref{fig-3} for the FCGG case. We first split the whole system into grids of small sizes, typically matching the particle dimension. Next, each lattice site is assigned an order-parameter value 
$\psi=+1$, if the local density (calculated with the particles within a cut-off radius) is larger than a pre-assigned cut-off $\rho_c=0.35$, else $-1$. 
This gives us a lattice-like configuration with Ising-like spins $\pm 1$. 
The original continuum snapshot and its coarse-grained (lattice) version, for the FCGG, are shown in Figs.~3(a) and (b), respectively.
We, in fact, implemented this method before calculating $C(r,t)$ for all the model systems.

\begin{figure}[!ht]
\centering
\includegraphics[width=9cm, clip]{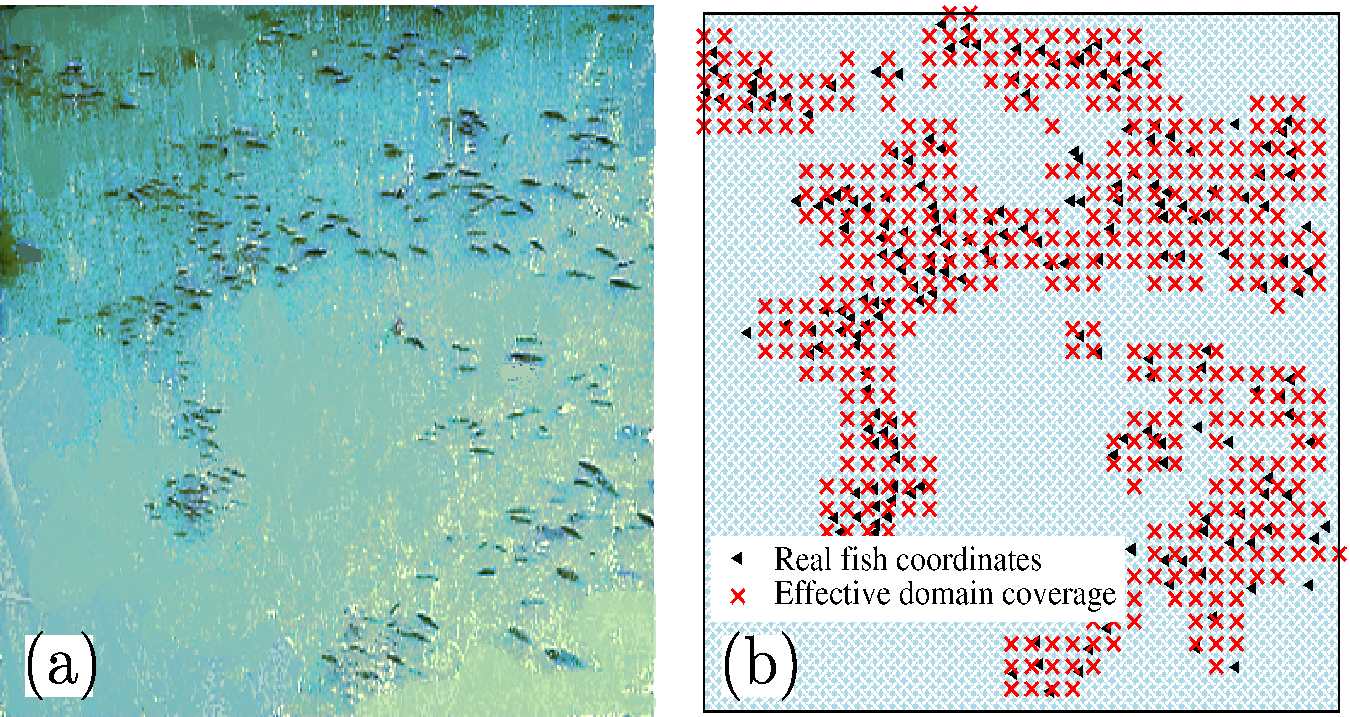}
\caption{(a) A photograph of a school of fish on the surface of a water body. (b) Black spots are off-lattice representations of the fish coordinates, identified using a plot digitizer \cite{digiplot}. The red marks correspond to coarse-grained representation. For the latter purpose, both the depth and length (having a ratio $1:5$, approximately) of a fish was considered, leading, on an average, to the (coarse-grained) area fraction same as the theoretical models.}
\label{fig-4}       
\end{figure}

\begin{figure}[!ht]
\centering
\includegraphics[width=9cm, clip]{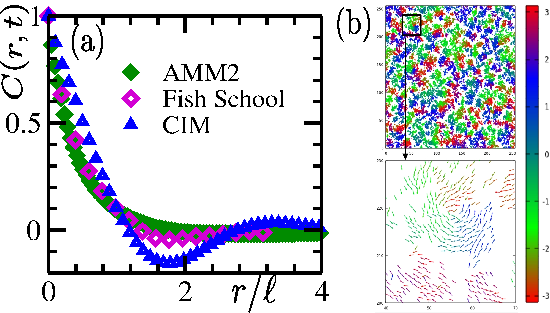}
\caption{(a) Plots of $C(r,t)$, versus $r/\ell$, for the school of fish, in Fig~\ref{fig-4}(a), are compared with those from the AMM$2$ and CIM. (b) Snapshot of AMM$2$ and demonstration of vortex formation in the velocity field. The color coding is for visual clarity of velocity directions. These results are for $f_A=10$.
}
\label{fig-5}       
\end{figure}

In Fig.~\ref{fig-4}(a) we show the photograph of a school of fish. 
We digitize the snapshot, using a software \cite{digiplot}, to pick the coordinates of each fish.  Furthermore, those points are mapped onto a square lattice of size $L=32$, and coarse-grained, following the description above, to realize the effective structural display. 
The original snapshot and the mapped spin-like (coarse-grained) configuration, have been shown in Figs.~\ref{fig-4}(a) and (b), respectively. 
There exists similarity with those for the models having velocity ordering feature.  
In order to quantify this we compare the $C(r,t)$ for the school and AMM2 in Fig.~\ref{fig-5}(a).  There exists a nice overlap of data. The agreement of $C(r,t)$ for the school of fish is  (somewhat) better with that for AMM2 than with those for FCGG and AMM1. Superiority of AMM2, over AMM1, perhaps is because of the (physical) volume exclusivity in AMM2. Here, we also plot the $C(r,t)$ corresponding to the Ising case, to emphasize the difference between the pattern formation in systems with and without velocity field. 
In the upper frame of Fig.~\ref{fig-5}(b) we have shown again a snapshot from AMM$2$. The color coding there is for clearer representation of the velocity directions of constituents. An enlarged portion is shown in the lower frame. This indicates vortex formation. Such structures can be seen in real active matters as well. In a future communication it will be interesting to compare the corresponding correlation function. 

\section{Conclusion}

We have considered a few model systems, viz., granular \cite{goldhirsch1993clustering} and active systems \cite{Vicsek1995,Das2017}, showing velocity alignment. 
Different microscopic evolution mechanisms bring inhomogeneity in the density field showing particle-depleted and particle-rich regions or clusters, within which the velocities are ordered. These density-field structures have been analysed via suitable order-parameter correlation function \cite{bray199341}, $C(r,t)$, and its Fourier transform, the structure factor \cite{bray199341}, $S(k,t)$. Both these quantities show nice matching over significant length-scales, depicting the similarity in the corresponding structures in different systems. 
We further find that there exists good agreement of the pattern formation in these models with that in a ``real'' active matter system, viz., a school of fish. 
While it requires further study to make a more concrete conclusion, the analyses above can be used to formulate a protocol to model naturally occurring assemblies accurately. For a system exhibiting clustering, one can compute structural quantities.  
Comparison of these within the candidate models, with the representative real ones, may lead to the choice that describes the complex natural system in a best quantitative manner. 

The authors thank N. Vadakkayil for a critical reading of the manuscript.

\bibliography{reference.bib}

\end{document}